\documentstyle [aaspp,epsf]{article}
\tighten
\begin{document}
\title {Probing Large Scale Structure using Percolation and Genus Curves}
\author {Varun Sahni \footnote{Inter-University Centre for 
Astronomy \& Astrophysics, Post Bag 4, Pune 411007, India;},
B.S. Sathyaprakash \footnote {Theoretical Astrophysics, 
California Institute of Technology, Pasadena, CA 91106 and
Department of Physics and Astronomy, UWCC, Cardiff, CF2 3YB, United Kingdom;}
and Sergei F. Shandarin \footnote {Department of Physics and Astronomy, 
University of Kansas, Lawrence, KS 66045.}}
\begin {abstract}
We study topological properties of large scale structure 
in a set of scale free N-body
simulations using the genus and 
percolation curves as topological characteristics. 
Our results show that as gravitational clustering advances, the
density field shows an increasingly pronounced departure from Gaussianity
reflected in the changing shape of the percolation curve and the
changing amplitude and shape of the genus curve.
Both genus and percolation curves differentiate
between the connectedness of overdense and underdense regions
if plotted against the density. When plotted against the filling factor 
the percolation curve alone retains this property.
The genus curve shows a pronounced decrease
in amplitude caused by phase correlations in the non-linear regime.
Both genus and percolation curves provide
complementary probes of large scale structure topology 
and can be used to discriminate
between models of structure formation and the analysis of observational
data such as
galaxy catalogs and MBR maps. 

\end {abstract}

\keywords {\noindent Subject Headings: (cosmology:) large-scale
structure of universe--galaxies: clusters: general--methods: numerical}

\section {Introduction}
It is reasonably well established that, far from being randomly
distributed, galaxies are strongly clustered and form a roughly cellular
three-dimensional structure consisting of filaments, sheets and clusters
separated by large voids
(\cite{delgh91}).
More than a decade ago Zel'dovich and Shandarin advocated using 
percolation theory to explore the 
topological properties of the large scale distribution of galaxies
(\cite{zel82}, \cite{sh83}, Shandarin \& Zel'dovich 1983, 1989). 
Percolation theory is also a useful
means with which to understand morphology viz.
the filamentarity/planarity/clumpiness of a distribution 
(\cite {sss96a}). 
A related topological approach complementary to percolation and 
involving the genus characteristic was suggested in
\cite{gmd86}. In the present study we, for the first time,
 apply techniques based
both on percolation analysis and the genus curve to 
the same set of N-body simulation data. 

It is well known that, as an initially random distribution of matter evolves
under self gravity, it rapidly develops non-Gaussian features 
(e.g. \cite{sc95}).
We assess the relative merits of the genus curve and the percolation curve
in providing measures of non-Gaussian features in such a distribution.
Both genus and percolation complement conventional 
indicators of clustering
by providing an estimate of the `connectedness' of
structure missed by standard estimators such as the two-point correlation 
function (which lacks phase information vital to an understanding of 
large scale coherence). 

To this end we have studied N-body simulations in an $\Omega=1$ universe.
We examined models with scale invariant initial spectra
$P_{in}(k) \equiv \left <
\left | \delta_k \right |^2 \right > \propto k^n,$ $n=-2$, $-1,$ $0$ and $1$.
The simulations are studied at several epochs,
each characterized by the scale of
nonlinearity $k_{\rm NL}^{-1}$ at that epoch measured in units of the fundamental
mode $k_f\equiv 2\pi/L,$ $L$ being the length of the simulation box.
For the sake of brevity we present results for
$n=-2$ and $n=0$ models, at epochs when the nonlinear scale is:
$k_{\rm NL}=64,$ 16 and 4. 
These spectral indices can be considered as the lower and upper limits
of the slope of the initial spectrum on the galaxy and supercluster scales.
N-body simulations are performed on a grid of size $128^3$ 
employing a particle-mesh algorithm (Melott \& Shandarin 1993).
Density fields are constructed on a reduced grid of
size 64$^3$ and subsequent studies of
percolation and genus are carried out on these fields.

\section {Percolation Analysis}
The starting point for percolation analysis is a rule or a criteria
by which to define structures. 
For instance, given the density field of matter in
the Universe and a density threshold, we can define a cluster 
(or a void) as a connected overdense (underdense)
region, connectivity being defined using a friends-of-friends 
or `nearest neighbors' algorithm 
\footnote {We use six nearest neighbors on the cubic grid 
to identify structures.
Although our results do depend on details of the nearest neighbors 
scheme, our main conclusions regarding the relative discriminating 
power of genus \& percolation curves do not.}.
The aim of percolation analysis is to study the connectedness of
structure as a function of the density threshold.
In an infinite medium varying the density threshold leads to a `percolation
transition' as the volume fraction in the largest cluster changes 
rapidly from almost zero to unity when the density
threshold crosses a critical value.
(In what follows we use density contrast 
$\delta =(\rho-\bar{\rho})/\bar{\rho}$, instead of density $\rho$,
to define the percolation transition.)

In reality one deals with finite systems and by an `infinite' structure 
one means a structure that spans the entire simulation box or observational 
sample. 
In this paper we will only consider finite systems and shall use the
term {\it simulation box} as a generic term which can refer to an observational
catalog or an N-body simulation box. 
For illustrative purposes we shall discuss the percolation of 
overdense regions keeping in mind that our arguments are equally valid for 
underdense regions (voids).
 
Gaussian random fields percolate at the critical filling factor
$FF_C \simeq 16\%$ regardless of the spectrum 
(filling factor -- henceforth $FF$ -- 
is the total volume in all clusters/voids above/below
the density contrast threshold divided by the simulation volume).
\footnote {$FF$ is the cumulative probability distribution function:
$FF = P(\delta > \delta_T)$.}
Density fields that have evolved under gravitational
instability typically percolate at lower levels of $FF_C$ 
depending upon the spectral index and the extent of non-linear evolution 
(\cite {ys96}). Similar conclusions can be made in the case of point like
distributions although the numbers are different and the natural reference
is the Poisson distribution ({\cite{ks93}).

\section {Effect of Dynamical evolution on Genus and Percolation Curves}

Traditional applications of percolation to gravitational clustering
focused primarily on the percolation threshold as a diagnostic measure
(\cite{sh83},\cite{k87},\cite{ds91}, \cite{ks93})
\footnote {The critical exponents remain almost unchanged as
clustering evolves
and are therefore poor indicators of non-Gaussianity (\cite {ks93}).}.
Although $FF_C$ (or $\delta_C$) 
is a useful characteristic of evolved
density fields, it has certain drawbacks
which make it unsatisfactory for the study of
a realistic distribution of galaxies. For instance, the
percolation threshold is sensitive to resolution, number of particles,
etc. In addition $FF_C$ is sensitive to the sample geometry 
and in cases when the observational
sample is wedge-shaped it becomes difficult to provide an unambiguous
definition of the percolation threshold (\cite{dw85}).

We investigate a powerful new statistic which does not face
most of the above limitations -- the {\it Percolation Curve}.
The percolation curve (henceforth PC) 
describes the {\it volume fraction} $v_{max}$ in the largest
structure (cluster/void) as a function of the density contrast 
threshold $\delta_T$ or $FF$. 
Formally, $v_{max}$  is defined as the ratio of the volume in
the largest cluster/void to the total volume in all clusters/voids
lying above/below a density contrast threshold: $v_{max}= V_{max}/FF$. 
($V_{max}$ is the fraction of the total volume in the largest cluster/void.)
 
In order to illustrate some salient features of the percolation curve
we first consider evolved density fields from
$n=-2$ and $n=0$ models at an epoch when the
scale of nonlinearity is $k_{\rm NL}= 64k_f, 16k_f, 4k_f.$ 
In Fig. \ref {fig-percolation} a,b we have plotted percolation curves
for clusters (thick solid lines) and 
voids (thick dashed lines) as a function of density contrast 
$\delta.$
Starting from a high density threshold (small $FF$)
we find
initially only one cluster in our field which corresponds to the highest
density peak, as a result $v_{max} = 1.$ However,
this is a finite system effect and we shall ignore it.
Lowering the threshold (increasing $FF$) we find several isolated clusters,
all of roughly comparable size so that the volume
fraction in the largest cluster is very small $v_{max} \ll 1$.
Lowering the density threshold still further results in the {\it merger} of clusters
leading to a rapid growth in $v_{max}$ and to the onset of percolation.
A further lowering of the threshold to very small values results in the
merger of almost all clusters so that $v_{max} \rightarrow 1$.
An identical procedure followed for underdense regions by gradually increasing
the density contrast threshold (increasing $FF$) results in a similar functional
form for the volume fraction in the largest void.
In our samples the largest cluster percolates between opposite faces of
the cube when its filling factor is about half the total $FF$. 
In most cases percolation also coincides
with the highest jump in the volume of the largest cluster which was used by \cite{delgh91}
as a working definition of the percolation threshold.
 
The solid/dashed vertical line in Fig. \ref {fig-percolation} represents the density
contrast threshold below/above which clusters/voids percolate.
We note that {\it both} clusters and voids percolate over a range of 
overlapping density contrasts --- a feature that is only possible in three 
or more dimensions --- and corresponds to what is commonly called a 
{\it sponge} topology for the density distribution. 

We clearly see that the percolation threshold $\delta_C$ (cluster) is higher
for $n=-2$ models relative to $n = 0$.
As the simulation evolves 
$\delta_C$ increases monotonically
for $n=-2$, as structures form and align on increasingly larger scales.
For $n=0$, $\delta_C$ initially 
increases but later {\it begins to decrease}
signaling the formation of small, isolated clumps.
It must be noted that for $n = 0$ the behavior of $\delta_C$ 
(cluster) is
non-monotonic whereas $FF_C$ (cluster) is monotonic (see Fig. \ref {fig-pcgc}), the reason for this
lies in the non-linear relationship between density threshold and filling
factor (\cite{ys96}). 
\footnote{The smallness of the filling factor in
simulations has been used to argue that the shape of overdense regions is 
likely to be filamentary (\cite{sss96a}). 
Note that error bars in Fig. \ref{fig-percolation} \& \ref{fig-pcgc} -- 
representing the dispersion computed with
the aid of four different realizations of a given power spectrum -- are very 
small.} From Fig. \ref {fig-percolation} we see that voids
find it easier to percolate as the simulation evolves, as a result
the range in densities when both phases percolate initially increases,
enhancing the extent of {\it sponge-like topology} in the distribution. 
We thus see that percolation analysis not only provides us with an appreciation
of different aspects of 
gravitational clustering but is sensitive enough to differentiate between
models with differing sets of initial conditions.

The percolation curve (PC) 
is in some respects similar to the `genus curve' (GC) which
can be formally expressed as an integral over the Gaussian curvature $K$
of the iso-density surfaces $S_{\nu}$ lying above/below a density threshold 
$\nu=\delta/\sigma_{\delta}$ by the Gauss-Bonnet theorem:
$4\pi G(\nu) = \int_{S_{\nu}} K d A$ (\cite{gmd86}).
\footnote{For discrete sets $4\pi G(\nu)
= - \sum_{i=1}^N D_i$, $D_i$ is the deficit angle at the $i$th vertex
of a polyhedral surface having $N$ vertices.} 
Multiply connected surfaces have $G \ge 0$ while simply connected
have $G < 0$.
For Gaussian Random fields the genus curve has a `bell shaped' form:
$G(\nu) = A (1 - \nu^2) \exp(-\nu^2/2)$ (\cite{hgw86,gwm87,gott89}).
(A corresponding analytical form for PC has so far eluded researchers.)
It should be borne in mind that realistic Gaussian fields are
defined over a {\it finite} interval, so that discreteness
effects must be taken into account when determining $G(\nu)$ {\it even for} 
the Gaussian case. 
Genus curves 
for models $n = -2, 0$ are shown as functions of $\delta$
in the right hand panels of 
Fig. \ref{fig-percolation}. We see that GC shows a rapid
departure from its original symmetric `bell shape' reflecting growth of 
non-Gaussianity
in the density distribution. Plotting GC against $\delta$ has one major
drawback: when plotted in this manner GC shows a 
difference
in topological properties of distributions related to each other by 
transformations $\delta \rightarrow f(\delta)$ (such as the log-normal)
when in fact no such difference exists.
A physically more relevant quantity to consider 
is the genus curve plotted against the
{\it filling factor}, which we discuss in the next section.

We would like to stress that, whereas properties of the {\it largest cluster} 
are used to determine PC, {\it all} clusters lying above/below a 
given threshold
are used to determine GC.
\section {Results, Discussion and Conclusions}

Percolation and genus curves are plotted as functions of $FF$
in Fig. \ref{fig-pcgc} for $n=-2$ (Fig. \ref{fig-pcgc}a) 
and $n=0$ (Fig. \ref{fig-pcgc}b), the scale of nonlinearity 
increasing from top to bottom ($k_{\rm NL}=64,$ 16 and 4).
For PC solid/dashed curves correspond
to the volume fraction in the largest cluster/void. 
Vertical solid/dashed lines correspond to the filling factor at percolation
$FF_C$ for overdense/underdense regions. 
The lightly dashed vertical line corresponds to $FF_C$ for 
Gaussianized fields constructed by randomizing the phases of the N-body 
particle distributions (\cite{ys96}).
For GC the solid line corresponds to overdense, and the dashed to
underdense regions. 

For a Gaussian distribution overdense and underdense regions
have identical topological properties. This fact is reflected in 
the uppermost panels of Fig. \ref{fig-percolation}b and \ref{fig-pcgc}b 
which show an early 
epoch of the $n=0$ model when the difference between the solid and 
dashed curves is small. In the $n=-2$ model the departure from Gaussianity
is evident even at this early stage.
Our method of plotting the genus curve is different
from convention which usually shows this curve plotted as a function of the
density contrast. Plotted in that manner the curve has a bell-shaped form
for Gaussian initial conditions.
Our method `folds the bell' and shows 
GC as a function of the filling factor. When plotted in this manner GC
appears to have a prescribed shape which it maintains {\it irrespective of the
evolutionary epoch} at which it is measured. The shape of GC appears to be
sensitive to the primordial spectral index: 
for small $n$ the separation between cluster
and void curves in GC is large, for large $n$ it is small 
(see Fig. \ref{fig-pcgc}).
The fact that the shape of the genus curve is very sensitive to the primordial
spectral index and shows little, if any, evolution leads us to feel
that it may be a useful indicator of the spectrum of primordial density 
fluctuations.

Since for spectra with substantial small scale power (such as $n = 0$), the 
asymmetry 
between clusters and voids in GC is small, one might
have expected the evolved genus curve to be a scaled version of GC for a
Gaussian random field. 
To check this we generated Gaussian fields (GC$_{\rm RAN}$) 
by randomizing the phases of 
the N-body particle distributions. 
Generating Gaussian fields in this manner allows some degree of control
over finite grid effects which now contribute equally to Gaussian and
non-Gaussian fields (\cite{ys96}).
Comparing GC with GC$_{\rm RAN}$ we found that
GC has a smaller amplitude than GC$_{\rm RAN}$ (see Table I).
Thus the decreasing amplitude of GC cannot be attributed to the 
(non-linear) evolution of the power spectrum. 
The difference in amplitude between GC and GC$_{\rm RAN}$ (more 
pronounced for spectra with 
small scale power) is caused by non-linear mode 
coupling and phase correlations
during advanced gravitational clustering.
In principle this effect could be used to probe the extent of non-linear 
evolution in the Universe by comparing red-shift data with 
numerical simulations 
at identical smoothing scales (see also \cite{mwg88,mel90}). 

Turning now to percolation, consider the percolation curves (PC) 
showing volume fraction in the largest cluster/void $v_{max}$ as a function 
of filling factor $FF$ (solid/dashed line) in Fig. \ref{fig-pcgc}.
The rapid growth in $v_{max}$ with increasing $FF$
indicates the onset of percolation. As the distribution evolves
the percolation threshold $FF_C$ becomes different for clusters and voids,
the difference: $FF_C(clusters) - FF_C(voids)$ increases with epoch
indicating the growth of non-Gaussianity.\footnote{We feel that the quantity
$FF_C(clusters) - FF_C(voids)$ may be less sensitive to sample geometry than 
$FF_C(clusters/voids)$.}
Divergence of the solid and dashed lines and the increase of the
area under the `hysteresis-like curve' in Fig. \ref{fig-pcgc} are also good
measures of the increase in non-Gaussianity. It is worth noting that the
vertical thin dashed line marking the percolation threshold in Gaussian
fields (constructed by randomizing phases) remains virtually unchanged.

The relative ease with which a distribution percolates relative to a Gaussian
has been used to categorize some of its topological properties.
For instance a distribution in
which overdense regions (clusters) percolate at lower $FF$ than 
Gaussian is said to
possess a `network' topology, whereas the reverse signals
a `meat-ball' topology. For voids, percolation at greater $FF$
than Gaussian, implies a bubble-topology consisting of isolated
underdense regions surrounded by overdense `cluster walls'.
Similar conclusions have also been drawn for the genus curve
(\cite{mel90}). 
Our results show easier percolation for overdense regions 
indicating a shift towards a network-like topology. 
Underdense regions on the other hand find it harder to
percolate and so have a bubble-like topology. 
\footnote{Note: The above definition of topology depends upon whether one
uses $FF$ or
$\delta$ to measure percolation. For instance $\delta_C$
{\it always decreases} for voids whereas $FF_C$ increases. 
Thus percolation of voids turns out to be either easier or more difficult
than Gaussian depending upon whether $\delta$ or $FF$
has been used to quantify percolation.}

Comparing GC and PC at identical epochs we find the degree of asymmetry
between clusters and voids to be much more pronounced for PC. 
As a result the increasing difference in topological properties of
overdense and underdense regions -- a clear indicator of non-Gaussianity --
appears to be better encapsulated in PC
than GC when both are plotted against the filling factor. 

The fact that the percolation curve is a sensitive diagnostic of
non-Gaussianity could be used
for the analysis of galaxy distributions 
and maps of the Cosmic Microwave Background (CMB). 
CMB maps derived from string/texture models of
structure formation are likely to be significantly non-Gaussian on small angular
scales and
in a future study we shall use percolation analysis to compare string/texture 
MBR 
maps with CDM maps (\cite{moess96}).

Finally, our results clearly imply that gravitational clustering
cannot be described by a mapping between initial and final states
such as: $\delta_{final} = f(\delta_{initial})$. Such a mapping 
would leave the percolation curve unaltered. 
This is in conflict with our results which unambiguously show that 
the percolation curve evolves during gravitational clustering.

\noindent {\bf Acknowledgments:}
Acknowledgments are due to the Smithsonian Institution, Washington,
USA, for International travel assistance under the ongoing Indo-US
exchange program at IUCAA, Pune, India.
We thank Adrian Melott for useful discussions, 
one of us (BSS) would like to thank him and 
the Department of Physics and Astronomy,
University of Kansas at Lawrence for hospitality where
this work was initiated. 
SS acknowledges
NSF grant AST-9021414, NASA grant NAGW-3832, and 
the University of Kansas
GRF-95 grant. BSS would like to thank Kip Thorne for hospitality
and encouragement and acknowledges 
Caltech NSF Grant AST-9417371.

\newpage
\begin {figure}
\caption {Percolation (left panels) and genus (right panels) curves 
are plotted as functions of the density contrast $\delta$ for 
scale free models of gravitational clustering (a) $n=-2$ and 
(b) $n=0.$ 
Solid and dashed curves in the left panels correspond to $v_{max}$ for
the largest cluster and void respectively. Vertical solid/dashed lines
mark the threshold describing percolation between opposite faces of the cube 
for clusters/voids respectively.} 
\label {fig-percolation}
\end {figure}

\begin {figure}
\caption {Percolation (left panels) and genus (right panels) curves 
are plotted as functions of the filling factor for N-body simulations of scale 
invariant spectra with index (a) $n = -2$ and  (b) $n=0$.
Panels illustrating percolation have solid/dashed lines showing $FF$
in the largest cluster/void ($v_{max}$) as a function of the total $FF$.
Heavy vertical solid/dashed lines mark the onset of percolation 
in the overdense/underdense phases; 
thin dashed lines mark percolation in Gaussian fields with
identical spectra.
Panels illustrating genus have solid/dashed lines showing the
genus curve for overdense/underdense regions plotted against
the total $FF$.} 

\label {fig-pcgc}
\end {figure}

\newpage
\def \tablerule {\noalign {\hrule}}
\begin {table}
\caption {Evolution of amplitude of the genus curve (at FF
$ = 0.5$) relative to Gaussianized fields
((GC$_{\rm RAN}$-GC)/GC$_{\rm RAN}$) is tabulated for $n=0$ and $-2$ models
at three expansion epochs.}
\vskip 0.2 true cm
\begin {tabular} {ccc}
\tablerule
$k_{\rm NL}$ & $n=0$ & $n=-2$ \\
\tablerule
64    & 0.00 & 0.33 \\
16    & 0.32 & 0.58 \\
 4    & 0.63 & 0.67 \\
\tablerule
\end {tabular}
\end {table}

\begin{thebibliography}{}
\bibitem [Dekel \& West 1985] 
{dw85}
Dekel A. \& West M.J. 1985, ApJ, 288, 411.

\bibitem [Dominik \& Shandarin 1992] 
{ds91}
Dominik K. \& Shandarin S.F. 1992, ApJ, 393, 450

\bibitem [Klypin 1987] 
{k87}
Klypin A.A. 1987, Soviet Astron., 31, 8

\bibitem [Klypin \& Shandarin 1993] 
{ks93}
Klypin A.A. \& Shandarin S.F. 1993, ApJ, 413, 48

\bibitem [de Lapparent, Geller, \& Huchra 1991]
{delgh91}
de Lapparent, V., Geller, M.J. \& Huchra, J.P. 1991, ApJ, 369, 273

\bibitem [Gott, Melott, \& Dikinson  1986]
{gmd86}
Gott, J.R.,  Melott, A.L. \& Dickinson, M. 1986, ApJ, 306, 341

\bibitem [Gott et al. 1987]
{gwm87}
Gott, J.R., Weinberg, D.H. \& Melott, A.L., 1987, ApJ, 319, 1.

\bibitem [Gott et al. 1989]
{gott89}
Gott, J.R. et al. 1989, ApJ, 340, 625.

\bibitem [Hamilton et al. 1986]
{hgw86}
Hamilton, A.J.S., Gott, J.R. \& Weinberg, D.H., 1986, ApJ, 309, 1.

\bibitem [Melott et al. 1988]
{mwg88}
Melott, A.L., Weinberg, D.H. \& Gott, J.R., 1988, ApJ, 328, 50.

\bibitem [Melott 1990]
{mel90}
Melott, A.L. 1990, Physics Reports, 193, 1

\bibitem [Melott \& Shandarin 1993]
{msh93}
Melott, A.L. \& Shandarin, S.F. 1993, ApJ, 410, 469

\bibitem [Moessner et al. 1996]
{moess96}
Moessner, R. et al.  1996, In preparation.

\bibitem [Sahni \& Coles 1995] 
{sc95}
Sahni V. \& Coles P., 1995, Physics Reports, 262, 1

\bibitem [Sathyaprakash, Sahni, \& Shandarin 1996a] 
{sss96a}
Sathyaprakash B.S., Sahni V. \& Shandarin S.F. 1996a, ApJ,  462, L5-L8.

\bibitem [Sathyaprakash, Sahni, \& Shandarin 1996b] 
{sss96b}
Sathyaprakash B.S., Sahni V. \& Shandarin S.F. 1996b, MNRAS, submitted

\bibitem [Shandarin 1983]
{sh83}
Shandarin, S.F. 1983,  Soviet Astron. Lett., 9, 104

\bibitem [Shandarin \& Zel'dovich 1983]
{shz83}
Shandarin, S.F. \& Zel'dovich Ya. B. 1983, Comments Astrophys., 10, 33

\bibitem [Shandarin \& Zel'dovich 1989]
{sz89}
Shandarin, S.F. \& Zel'dovich Ya. B. 1989, Rev. Mod. Phys., 61, 185

\bibitem [Yess \& Shandarin 1996]
{ys96}
Yess C. \& Shandarin S.F. 1996, ApJ, 465, 2

\bibitem [Zel'dovich 1982] 
{zel82}
Zel'dovich Ya. B. 1982, Soviet Astron. Lett., 8, 102

\end{thebibliography}
\end {document}
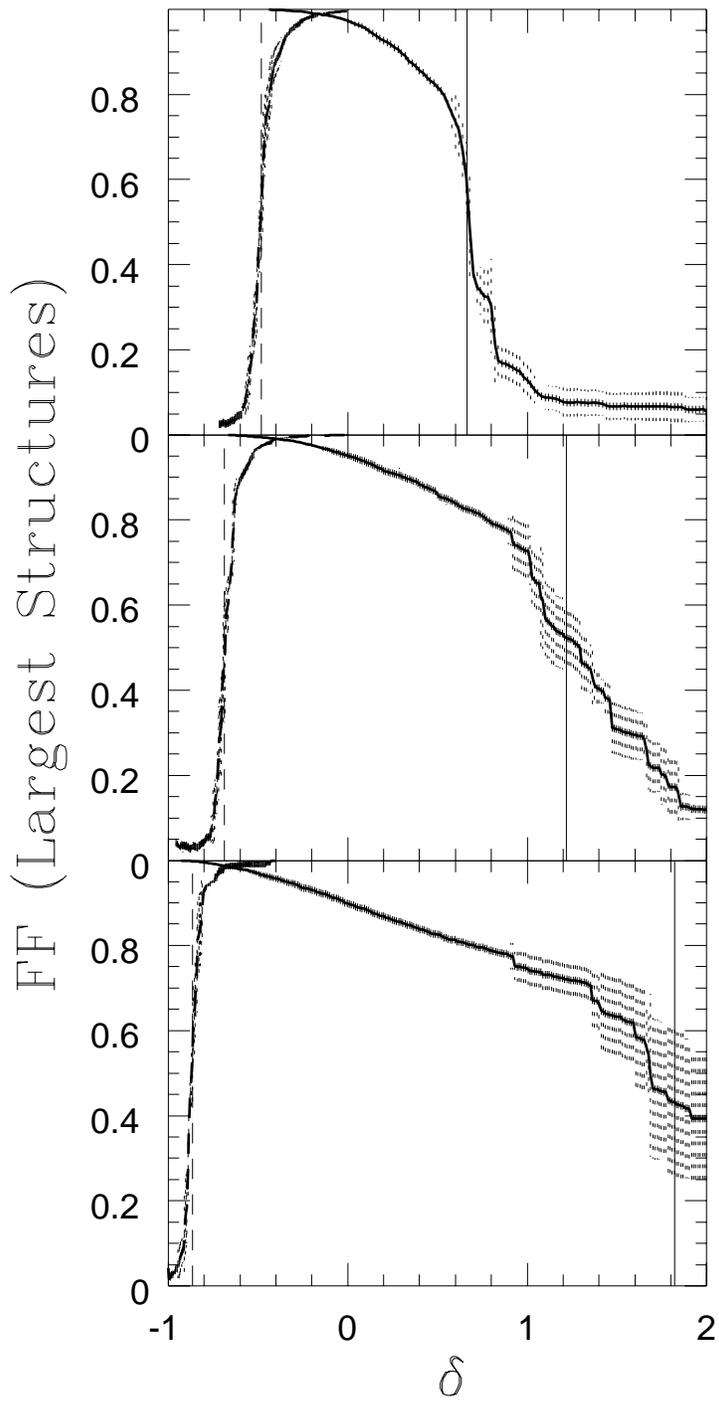
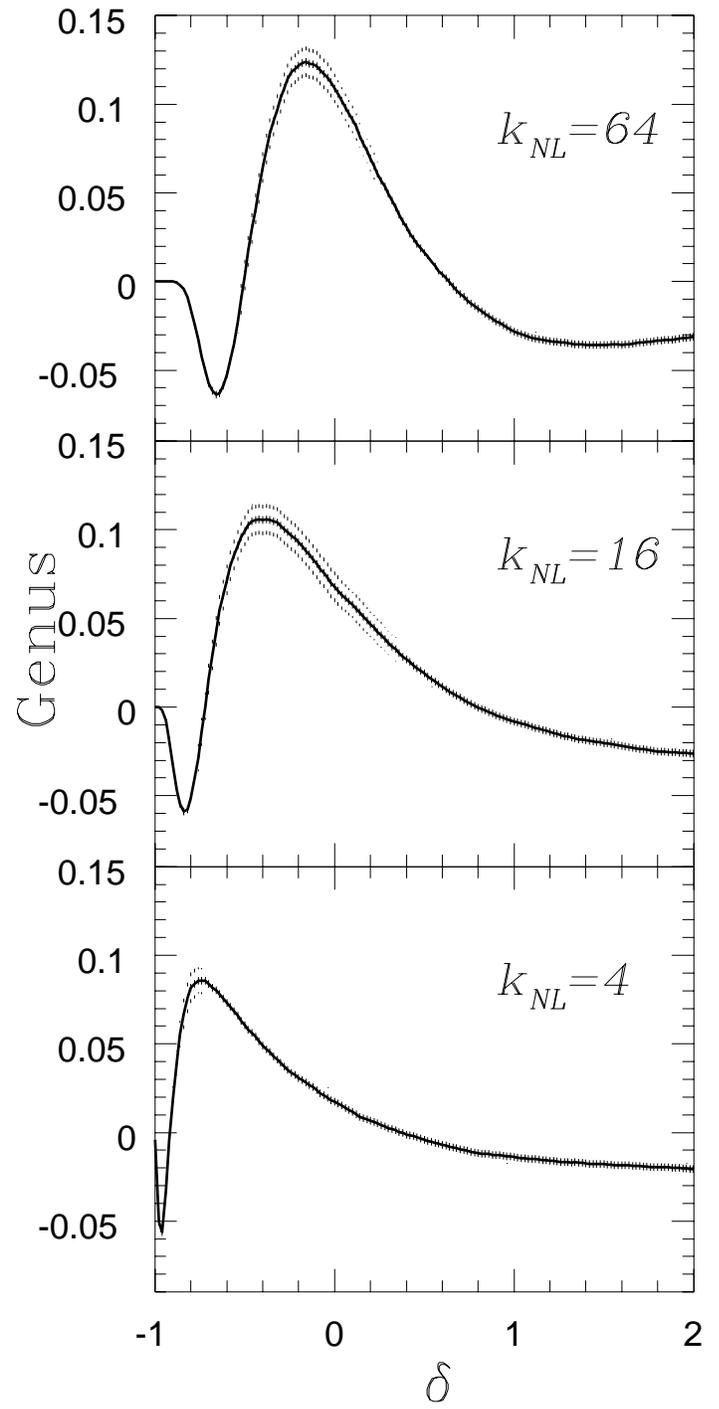
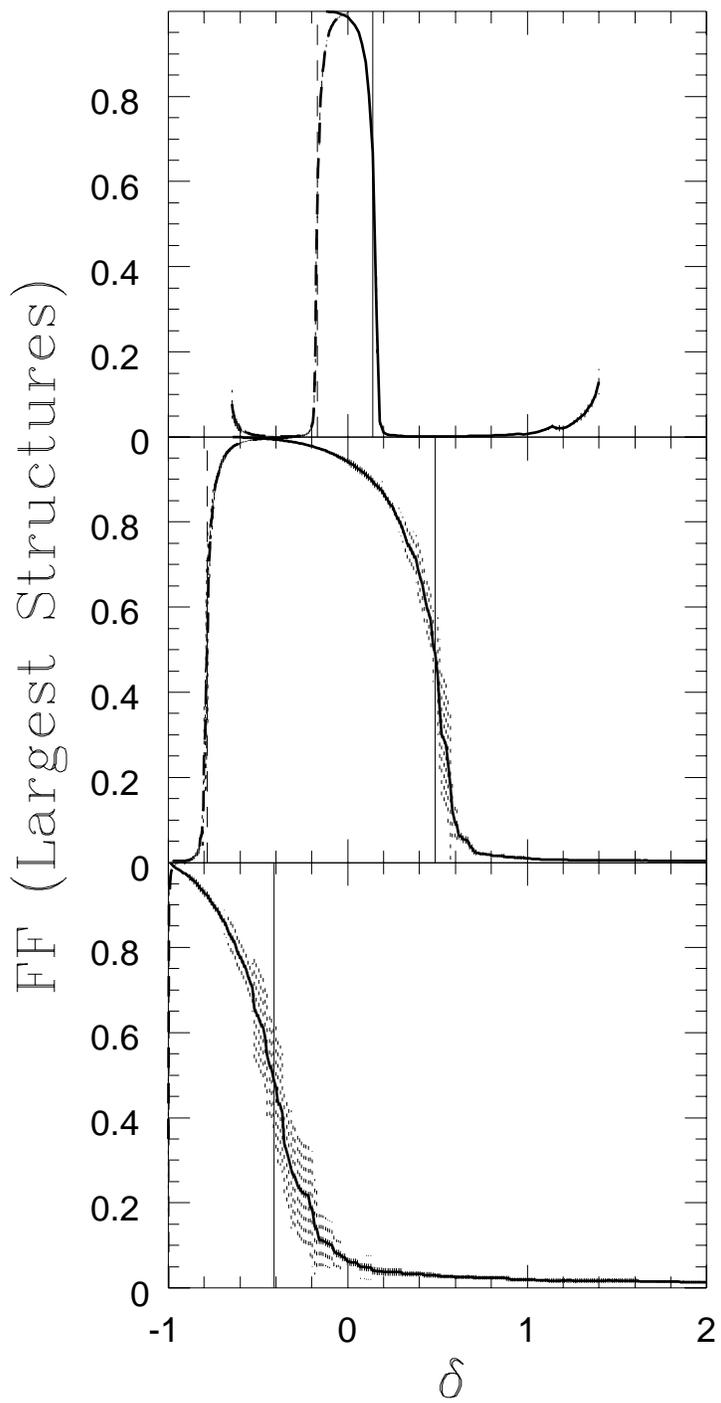
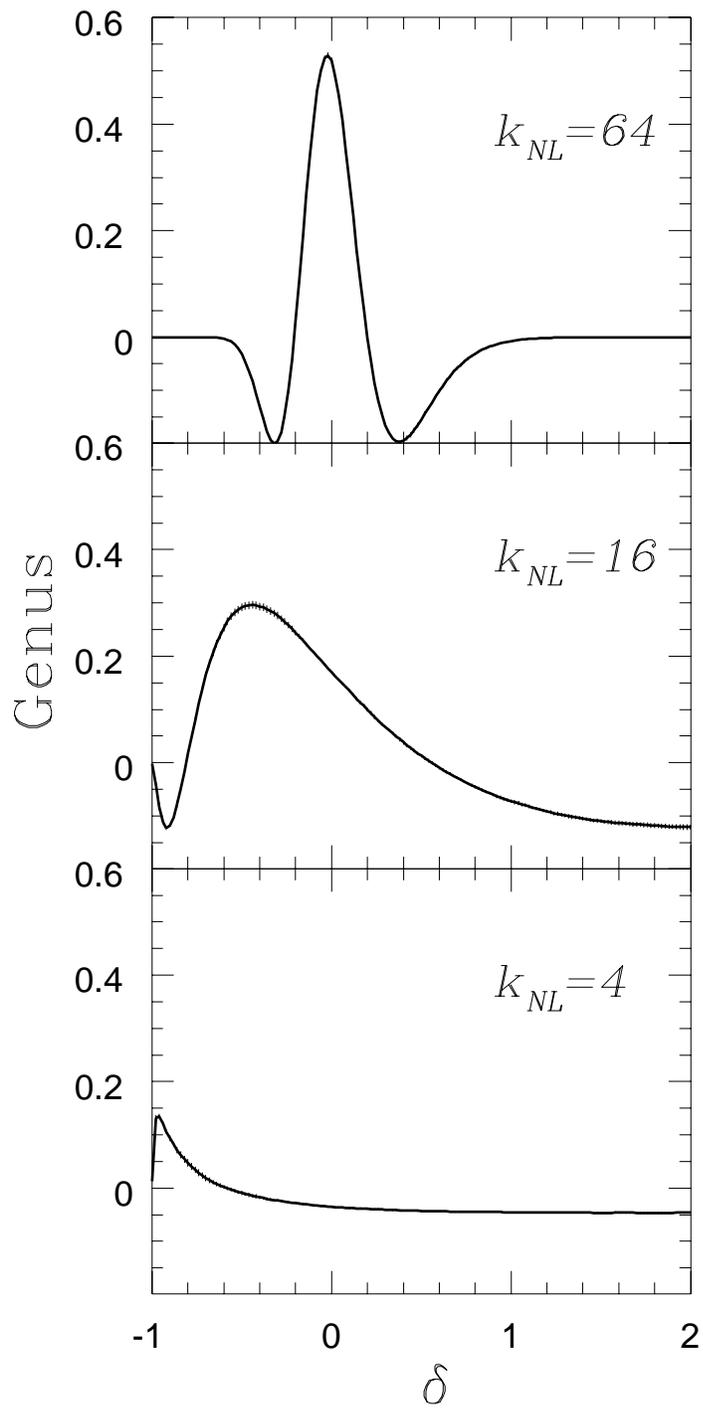
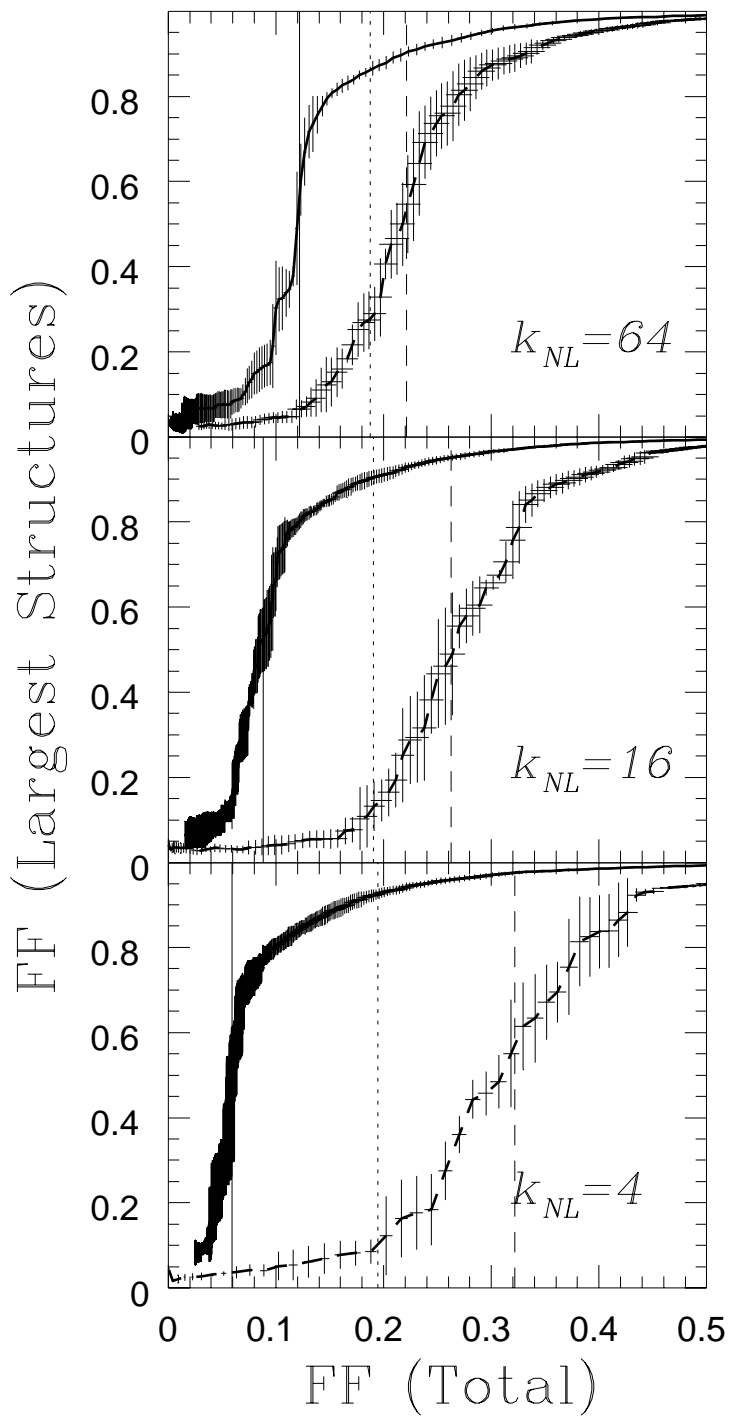
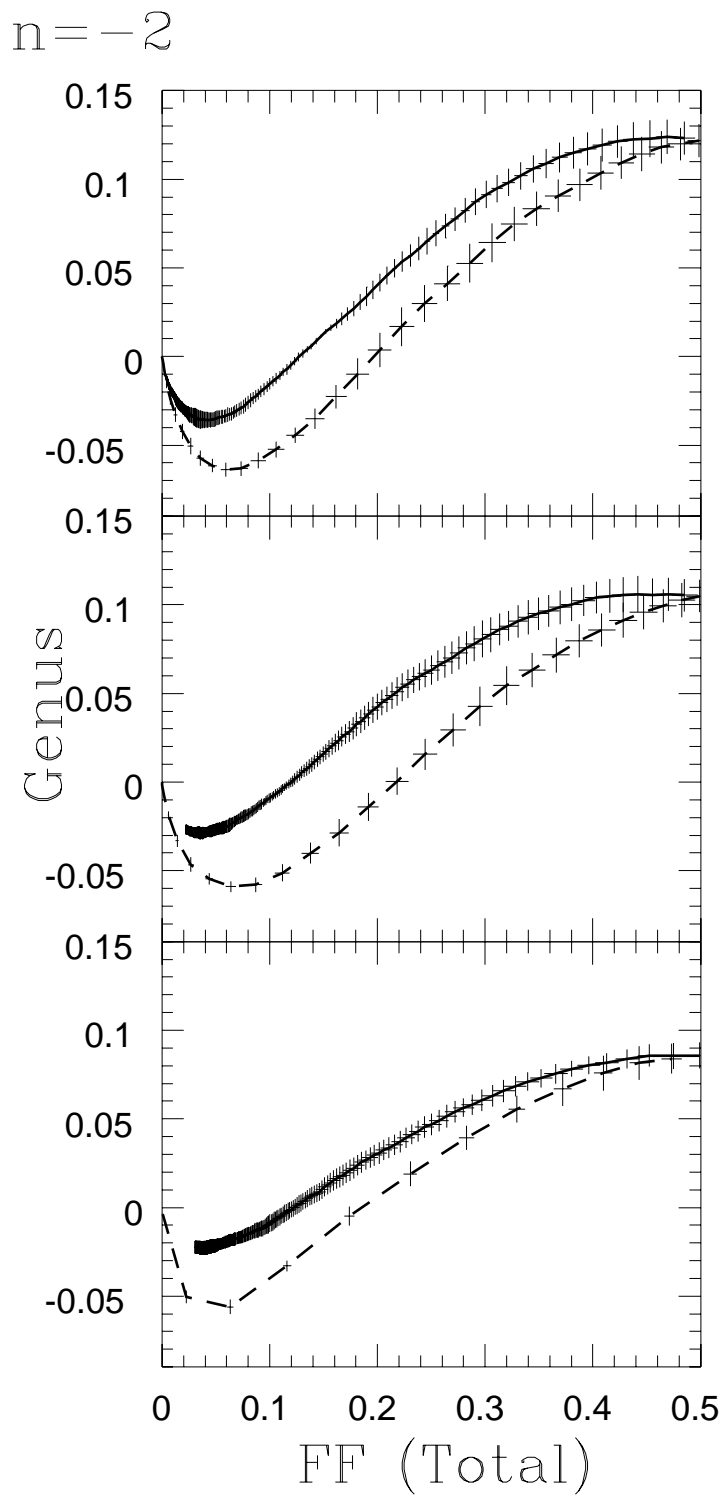
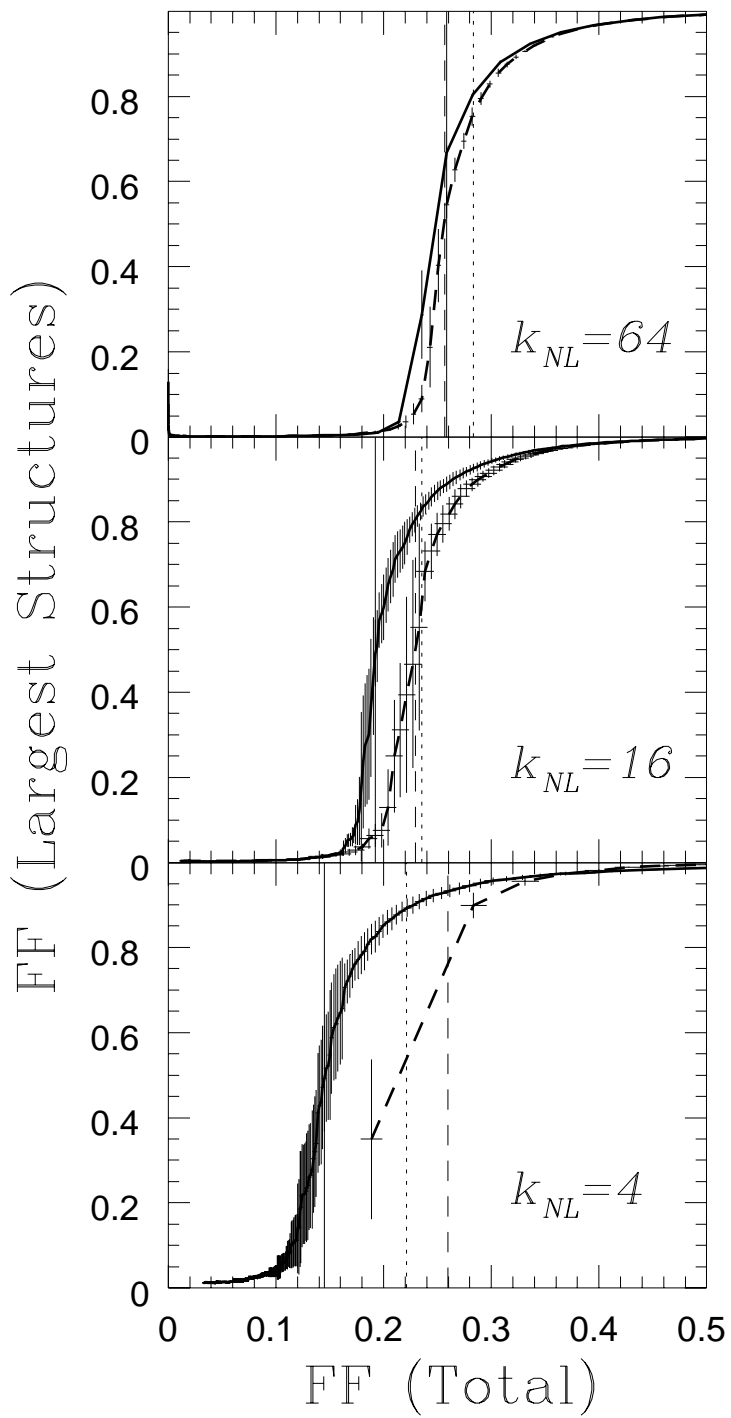
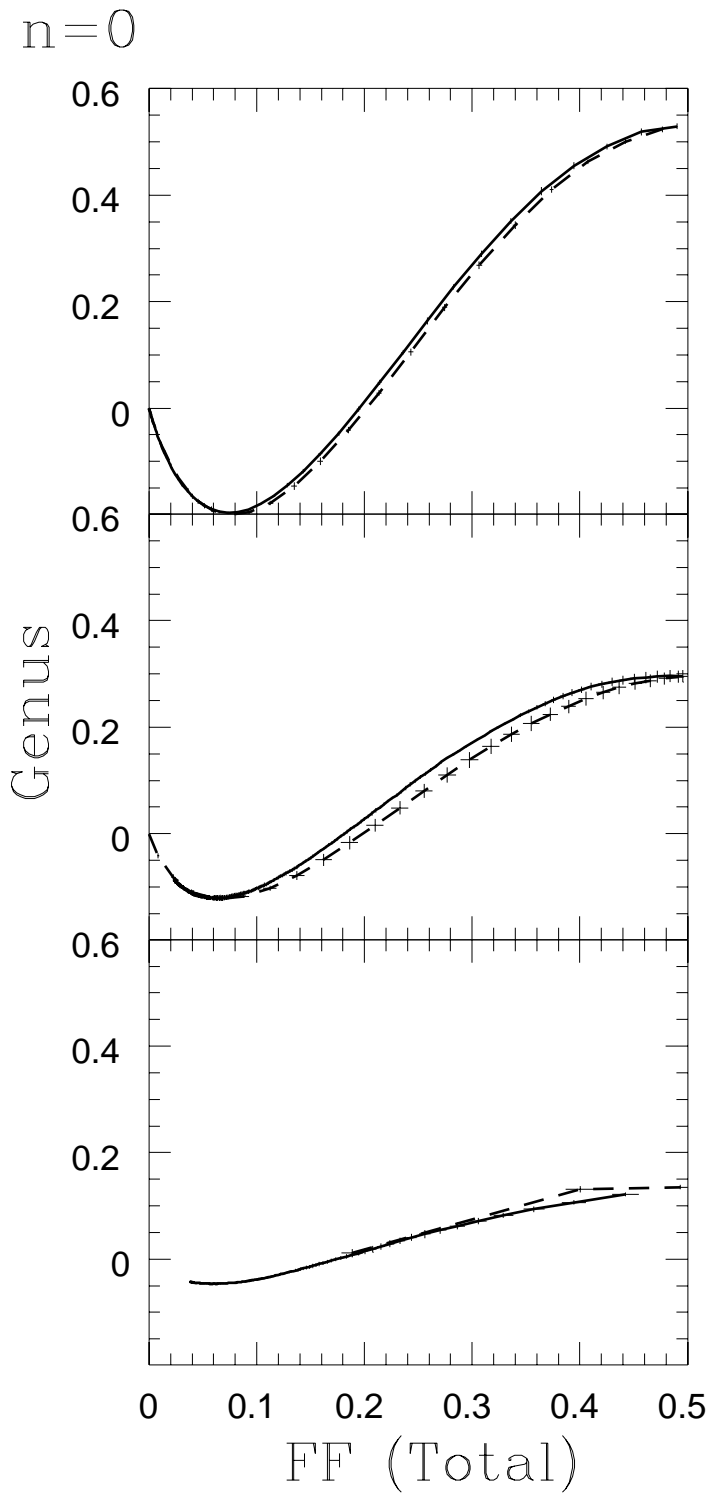
